\documentclass[prl,amssymb,superscriptaddress, twocolumn]{revtex4}

\usepackage{amsmath}
\usepackage{amssymb}
\usepackage{subfigure}
\usepackage{graphicx}
\usepackage{ulem}
\usepackage{multirow}
\usepackage{enumitem}
\usepackage{lipsum, babel}

\newcommand{\be}{\begin{equation}}
	\newcommand{\ee}{\end{equation}}

\newcommand{\bea}{\begin{eqnarray}}
	\newcommand{\eea}{\end{eqnarray}}

\usepackage{xcolor}

\begin{document}

\title{Infinite symmetry prevents disorder-induced localization in 2D}

	\author{C.\,A.\,Trugenberger}
	\affiliation{SwissScientific Technologies SA, rue du Rhone 59, CH-1204 Geneva, Switzerland}
	\affiliation{Division of Science, New York University Abu Dhabi, Abu Dhabi, United Arab Emirates}
	
	
	%
\begin{abstract}
We show that 2D gapped many-body quantum states are constrained by an infinite-dimensional symmetry which renders them transparent to weak disorder. This prevents disorder-induced localization when interactions are strong enough to open a gap. Using purely algebraic methods we derive all possible quantum states near the superconductor-to-insulator (SIT) transition and we compute the meson spectrum of superinsulators. 
		
\end{abstract}
	
\maketitle

\section{Introduction}

Ever since Anderson's seminal paper \cite{anderson} showing that disorder can localize single particle states, it has been an open question if localization can survive strong interactions in quantum many-body states. In \cite{basko} it was shown that this is the case in perturbation theory, provided all single-particle states are localized and this has been associated with finite-temperature perfectly insulating behaviour, even in higher dimensions, such as 2D films \cite{altshuler, mblrev2}. 

However, it has by now been shown that this perfectly insulating behaviour at finite temperatures is due to superinsulation \cite{dst, vinokur, dtv} (for a review see \cite{enc}), i.e. instanton-driven electric confinement, and not to disorder-induced localization. On one side, the perfectly insulating behaviour takes place also in regular, perfectly ordered systems \cite{natcomm}, thereby excluding disorder as its root cause; on the other side, the confining linear potential has been directly measured in experiment \cite{dynamic}. As explained in \cite{natcomm}, the superinsulating behaviour is sometimes mistaken for a Bose Mott insulator in systems which are too small, so that only the microscopic degrees of freedom are observed instead of the collective infrared behaviour of the whole many-body quantum state. 

Here we illustrate why, in 2D, weak disorder cannot have any effect on the nature of 2D (fully) gapped many-body ground states and their excitations, where by weak disorder we mean random perturbations whose energy scale is smaller than the gap and which do not alter the nature of the microscopic degrees of freedom, e.g. they do not break Cooper pairs. Note that, in 2D superconductors, the gap for vortex deconfinement is typically smaller than the Cooper pair binding energy. We show that interactions strong enough to open a gap in 2D prevent disorder-induced localization by the constraints of an infinite-dimensional dynamical (or spectrum-generating) symmetry algebra. Such strong interactions render the quantum many-body states transparent to disorder; as in the quantum Hall liquids, disorder can only pin the excitations, helping, e.g., to broaden the resistance plateaus (although, strictly speaking, even here disorder is not really necessary \cite{kivelson}) but has no influence on the nature of these excitations and of the ground state. For superconducting states, this is a generalization of Anderson's theorem: the variation of the superconducting parameters depends only on interactions, not on disorder.  For superinsulating systems, the resistance at low, but finite temperatures is infinite because the excitation spectrum does not include charged states due to the strong interactions and thus pinning, even if present, is irrelevant. This is the reason why this phenomenon occurs also in ordered systems. 

The infinite symmetry implies an infinite number of constraints on many-body quantum states, which make them robust against perturbations, and an infinite number of quasi-local constants of motion. For chiral systems, like the quantum Hall liquids, this has been extensively discussed in \cite{ctz2, ctz1, ctz3, karabali1, karabali2, flohr}. Here we extend these results to non-chiral, parity-invariant systems, also ones, as superinsulators, which are not topologically protected. In particular, we will use the infinite symmetry to derive, by purely algebraic methods, the entire meson spectrum of superinsulators. These results show that the protection by an infinite symmetry algebra is even more fundamental than topological order. 

Finally, we would like to conclude this introduction by stressing that this is a distinct mechanism from Hilbert space fragmentation (see \cite{moudgalya} for a recent review). Symmetries imply a number of Hilbert space disconnected sectors growing polynomially with system size, contrary to the exponential growth in Hilbert space fragmentation \cite{motrunich, sen}.

When dealing with a very large number of degrees of freedom it sometimes pays to look at the system as a whole, instead of trying to deduce its properties from the interactions of its components. It might happen that the entire system has a symmetry that helps identify the ground state and the full excitation spectrum by algebraic methods, without ever solving any dynamic equation for a specific Hamiltonian. This is the case, e.g. for the strong interactions of particle physics, which have a dynamical symmetry (albeit an approximate one) under the $SU(3)$ flavour group (for a review see \cite{georgi}) or for 2D statistical mechanics models at a second-order phase transition, which have a conformal symmetry (for a review see \cite{francesco}). 

\section{Symmetries}

When a dynamical symmetry is present, the spectrum of possible quantum states decomposes into irreducible representations (multiplets) of the symmetry group and the Hamiltonian can be expressed as a sum of Casimir operators of the symmetry algebra. As a consequence, the Hamiltonian is block-diagonal, each multiplet being characterized by the same energy. In the $SU(3)$ model of strong interactions, particles in each hadron multiplet have (approximately) the same mass. Each state in a multiplet is characterized by the conserved eigenvalues of the operators in the Cartan subalgebra. In the $SU(3)$ model, these are the hadron quantum numbers, called isospin and hypercharge. Particle multiplets can be composed by the fusion rules of the symmetry algebra, which are expressed, in this example, by the Clebsch-Gordon coefficients of $SU(3)$. 

In the case of the conformal symmetry there is one more complication: the symmetry algebra is infinite-dimensional. Therefore, the representations are also infinite-dimensional and the fusion rules have to be modified accordingly \cite{francesco}. Moreover, also the Cartan subalgebra becomes infinite-dimensional and, therefore, each representation is characterized by an infinity of conserved quantum numbers.  The infinite dimensionality, however also brings an advantage. Contrary to the $SU(3)$ case, for specific values of the central charge (see below) and the weights, there exist so called ``minimal models", representing sets of degenerate representations that are closed under the fusion rules. These degenerate representations have less states than generic ones since some of the original states have to be projected out to make the representations irreducible. These minimal models encode thus particularly robust theories with a minimal set of excitations: it is these minimal conformal models that classify critical points in 2D.  

Gapped many-body states are essentially incompressible at temperatures $k_{\rm B}T$ much lower than the gap $\Delta$, since all bulk excitations, including density fluctuations, are suppressed by a Boltzmann factor ${\rm exp}(-\Delta /k_{\rm B}T)$. In 2D, at the classical level, the configuration space of incompressible systems is generated from an initial configuration, say, a disk, by {\it area-preserving diffeomorphisms}. These can be represented as canonical transformations of a 2D phase space, described by complex coordinates $z$ and $\bar z$ (we normalize space coordinates by the typical length scale in the problem so that $z$ and $\bar z$ are dimensionless) and a Poisson bracket 
\begin{equation}
\left\{ f, g\right\} = \pm i \left( \partial f \bar \partial g- \bar \partial f \partial g \right) \ ,
\label{poisson}
\end{equation}
where the two signs correspond to the two possible polarization choices in which either $\sqrt{2} y$ is a ``coordinate" and $\sqrt{2} x$ a ``momentum" or the other way around. Infinitesimal area-preserving diffeomorphisms $\delta z = \{ {\cal L}, z \}$ and $\delta \bar z = \{ {\cal L},\bar z \}$ are given in terms of generating functions ${\cal L} (z, \bar z)$ which can be expanded in the basis ${\cal L}_{n,m} = z^n \bar z^m$ on ${\mathbb C} \backslash \{ 0 \}$. These basis generators satisfy the two infinite-dimensional algebras
\begin{equation}
\left\{ {\cal L}_{n,m}, {\cal L}_{k,l} \right\} = \mp i (mk-nl) \ {\cal L}_{n+k-1,m+l-1} \ ,
\label{classical}
\end{equation}
depending on which polarization was chosen in (\ref{poisson}). A 2D parity transformation interchanges $z$ and $\bar z$ and, correspondingly, the two algebras: if ${\cal L}_{n,m}$ satisfy the algebra with the upper sign, ${\bar {\cal L}}_{n,m}$ satisfy the algebra with the lower sign. Note that generators ${\cal L}_{n,m}$ with $n>0$ and $m>0$ form two closed sub-algebras. We shall restrict to these two subalgebras, since this is anyway required at the quantum level. We consider thus only generators ${\cal L}_{i-n,i}= |z|^{2i}z^{-n}$ and ${\bar {\cal L}}_{i-n,i} = {\cal L}_{i,i-n}=|z|^{2i}{\bar z}^{-n}$, where $i>0$ and where the upper polarization applies to ${\cal L}$ and the lower one to $\bar{\cal L}$. These two subsets of generators form the classical algebra of area-preserving diffeomorphisms, also called $w_{\infty}$ \cite{shen}.

As 2D classical incompressible configurations are generated by area-preserving diffeomorphisms, 2D gapped quantum many-body ground states must be irreducible representations of the quantum versions of the algebras (\ref{classical}), which are, of course, also infinite-dimensional. These can be obtained by the standard procedure of promoting Poisson brackets to commutators, $i\{,\} \to [,]$ (we use natural units). As usual, quantization entails replacing one of the coordinates $z$ or $\bar z$ with the derivative with respect to the other one. Since negative powers of a derivative do not exist, we must restrict the generators to the subsets ${\cal L}_{i-n,i}$ and ${\bar {\cal L}}_{i-n,i}$ defined above. These become the quantum operators ${\cal L}_{i-n,i} \to V^i_n = : z^{i-n} \partial ^i: $ and
${\bar {\cal L}}^{i-n,i} \to {\bar V}^i_n = : {\bar z}^{i-n}\bar \partial :$, where $\partial = \partial/\partial z$, $\bar \partial = \partial/\partial \bar z$ and $:\ \ :$ denotes Weyl ordering \cite{largeN}. These generators span the two commuting (on ${\mathbb C} \backslash \{ 0 \}$) chiral sectors $W_{1+\infty}$ and ${\overline W}_{1+\infty}$ of the algebra $W_{1+\infty} \otimes {\overline W}_{1+\infty}$ of quantum area-preserving diffeomorphisms. The name $W_{1+\infty}$ encodes the quantum character of the algebra by a capital ``W", while the explicit ``1" in the name indicates that the operators with $i=0$ are included in the algebra (see, e.g. \cite{frenkel}). 

The quantum algebra $W_{1+\infty}$ is given by
\begin{eqnarray}
\left[ V^i_n, V^j_m \right] &&= (jn-im) V^{i+j-1}_{n+m} + q(i,j,n,m) V^{i+j-3}_{n+m} + \dots 
\nonumber \\
&&+ \delta^{ij} \delta_{n+m} \ c \ d(i,n) \ ,
\label{walgebra}
\end{eqnarray}
where the structure constants $q$ and $d$ are polynomials of their arguments and the dots indicate similar terms involving the operators $V^{i+j-(2k+1)}_{n+m}$, whose exact form is somewhat cumbersome but not of importance for what follows. The first term on the r.h.s of (\ref{walgebra}) is the classical term, corresponding to (\ref{classical}); the following terms are quantum operator corrections. Finally, the last, c-number term is a quantum anomaly with central charge $c$, whose possible values will be explained below. The corresponding algebra ${\overline W}_{1+\infty}$ is the parity transform of $W_{1+\infty}$: once we know the quantum numbers of $W_{1+\infty}$ representations we can obtain those of ${\overline W}_{1+\infty}$ representations simply by reversing the sign of all parity-odd quantities. 

Notwithstanding the apparent complexity of the infinite-dimensional algebra $W_{1+\infty}$, its representation theory has been completely worked out \cite{kac1, kac2}. Moreover, also its minimal models have been derived \cite{ctz1} and these are in one-to-one correspondence with the Jain hierarchy \cite{jain} of incompressible quantum Hall fluids \cite{ctz1}, although new neutral modes with non-Abelian statistics at levels larger than one in the hierarchy are predicted. The robustness of the quantum Hall states and the precision of their fractional quantum numbers is traced back to the constraints of an infinite-dimensional dynamical symmetry \cite{ctz2, karabali1, karabali2, flohr}. 

\section{The $W^V_{1+\infty} \otimes W^A_{1+\infty}$ minimal models}

Here we shall extend these results to all 2D gapped states, including non-chiral ones, which provides a generic algebraic characterization of such states which is stronger than topological order. In particular, we shall focus on the consequences of the infinite dynamical symmetry $W_{1+\infty} \otimes {\overline W}_{1+\infty}$ for parity-invariant 2D gapped quantum many-body states and we will use these insights to classify the possible phases and their excitations near the superconductor-to-insulator transition (SIT) (for a review see \cite{goldmanrev}). In particular, we will identify possible 2D gapped many-body quantum states of matter with minimal models of the dynamical symmetry $W_{1+\infty} \otimes {\overline W}_{1+\infty}$.

We start discussing the algebra $W_{1+\infty}$; as previously mentioned we can obtain the corresponding results for ${\overline W}_{1+\infty}$ simply by a parity transformation. The generators $V^i_n$ are characterized by an integer conformal spin $i+1 \ge 1$ and an angular momentum mode index $n$, $-\infty < n <+ \infty$. The operators $V^0_n$ are the charge operators and satisfy the Abelian Kac-Moody algebra $\widehat U(1)$, which is the extension of $U(1)$ by a c-number central charge $c$ (see, e.g. \cite{francesco}) . The operators $V^1_n$ generate conformal transformations and satisfy the Virasoro algebra with central charge $c$ (see, e.g. \cite{francesco}). The operators $V^i_0$ form the infinite-dimensional Cartan sub-algebra. 

Irreducible, unitary, quasi-finite highest-weight representations of $W_{1+\infty}$ exist only if the central charge is a positive integer, $c=m \in {\mathbb Z}_+$. They are characterized by an $m$-dimensional weight vector ${\vec r}$ with real elements and are built on top of a highest weight state $|{\vec r}\rangle$ which satisfies \cite{kac1, kac2}
\begin{equation}
V^i_n |{\vec r}\rangle = 0 \ , \ \ \  \forall n>0, i\ge 0 \ ,
\label{hws}
\end{equation}
and is an eigenstate of the Cartan charges 
\begin{equation}
V^i_0 |{\vec r}\rangle = \sum_{n=1}^m m^i\left( r_n \right) |{\vec r}\rangle  \ ,
\label{cartan}
\end{equation}
where $m^i\left( r_n\right) $ are $i$-th order polynomials of the weight components. For example, the charge number $V^0_0 = q$ and scaling dimension $V^1_0 = h$ are given by
\begin{eqnarray}
q &&= \sum_{n=1}^m m^0 \left( r_n \right) = r_1 + \dots + r_m \ ,
\nonumber \\
h &&= \sum_{n=1}^m m^1 \left( r_n \right) = {1\over 2} \left[ \left( r_1 \right)^2+ \dots + \left( r_m\right) ^2 \right]  \ .
\label{chargescaling}
\end{eqnarray}
Each such highest-weight representation describes a (chiral) incompressible quantum state, subject to an infinity of constraints (\ref{hws}) and describing the ground state if all $r_n=0$ or a quasi-particle (quasi-hole) at the origin, characterized by an infinity of quasi-local charges (\ref{cartan}), otherwise. The infinite towers of states in the representations comprise the possible edge excitations. Due to incompressibility, the bulk Hamiltonian must be diagonal in the space of these representations. As a consequence, for temperatures and external excitations much smaller than the gap the quasi-local charges are conserved. Moreover, the infinite constraints render the states robust with respect to disorder. 

There exist two types of such irreducible highest-weight representations \cite{kac1, kac2}: generic representations if the weight vector ${\vec r}$ satisfies $(r_i-r_j) \notin {\mathbb Z}$, $\forall i\ne j$ and degenerate representations, which have $(r_i-r_j) \in {\mathbb Z}$ for some $i\ne j$. Generic representations are equivalent to the $\widehat U(1)^{\otimes m} $ representations with the same weight vector. Degenerate representations are one-to-one equivalent to those of the $\widehat U(1) \otimes {\cal W}_m$ minimal models in the limit $c\to m-1$ \cite{ctz1}, where ${\cal W}_m$ is the Fateev-Lykanov-Zamolodchikov algebra \cite{fateev1, fateev2}. This means that one-class degenerate representations at level $m$ contain an additive charged excitation and $(m-1)$ neutral modes whose weights add up modulo an integer combination of the simple roots of $SU(m)$. Note, however that contrary to the standard $SU(m)$ multiplets, these are infinite-dimensional representations: they are made by the $SU(m)$ highest weight, representing a neutral bulk excitation, and an infinite tower of edge excitations \cite{ctz1, ctz2}. The neutral bulk excitations combine according to the $SU(m)$ fusion rules and have, thus non-Abelian statistics.

The $W_{1+\infty}$ minimal models \cite{ctz1} are made by sets of such one-class degenerate representations ${\cal R}_{m, r, {\bf n}}$ that are closed under the fusion rules for making composite states. These sets can be obtained by choosing the weigths ${\vec r} = \sum_{i=1}^m n_i {\vec v}_i$ on a lattice that contains the $(m-1)$ simple roots of $SU(m)$. Such lattices are spanned by the basis 
\begin{equation}
\left( {\vec v}_i \right)_j = \delta_{ij} + r \ C_{ij} \ ,
\label{latt}
\end{equation}
where $r \in {\mathbb R}$ is a free parameter, $C_{ij} = 1$, $\forall i,j = 1, \dots, m$ and the excitations labels $n_i$ must be chosen to satisfy $n_1 \ge n_2 \ge \dots \ge n_m$ in order to avoid double counting. The charge number and scaling dimension of a bulk excitation with label ${\bf n}$ is given by
\begin{eqnarray}
q &&= \ \ \ {\bf t}^{\rm T} \cdot M \cdot {\bf n} \ ,
\nonumber \\
h &&={1\over 2} \ {\bf n}^{\rm T} \cdot M \cdot {\bf n} \ ,
\label{qchiral}
\end{eqnarray}
where ${\bf t} = (1, \dots, 1)$ and the matrix $M$ is the metric of the lattice,
\begin{equation}
M_{ij} = {\vec v}_i \cdot {\vec v}_j = \delta_{ij} + \lambda \ C_{ij} \ , \qquad \lambda = mr^2 +2r \ .
\label{metric} 
\end{equation}

In the following we shall focus on parity-invariant minimal models, characterized by the values of an integer central charge $m$ and a real number $r$ and defined as 
\begin{equation} 
{\cal M}_{\rm m, r}  = \bigoplus_{{\bf n}, {\bar {\bf n}}} {\cal R}_{m, r, {\bf n}} \otimes {\bar {\cal R}}_{m, r, {\bar {\bf n}}} \ ,
\label{pmodels}
\end{equation}
where the excitations labels ${\bf n}$ and ${\bar {\bf n}}$ lie on the same integer lattice and the quantum numbers are given by
\begin{eqnarray}
Q &&= q+ \bar q \ , \qquad \Phi = q- \bar q \ ,
\nonumber \\
H &&= h + \bar h \ , \qquad S = h -\bar h  \ ,
\label{qns}
\end{eqnarray} 
with $Q$ the charge number, $\Phi$ the vortex number, $H$ the scaling dimension and $S$ the spin. Actually, we will be interested specifically in examples which can be formulated as $W^V_{1+\infty} \otimes W^A_{1+\infty}$ where $V$ and $A$ denote ``vector" and ``axial-vector" components, so that every excitation is a combination of an elementary charge and an elementary vortex. 

To this end let us focus first on the simplest case $m=1$. In this case, we can obtain $W^V_{1+\infty} \otimes W^A_{1+\infty}$ minimal models by restricting to the ${\mathbb Z}^2$ sub-lattice for which $n\pm \bar n = {\pm \ \rm even \ integer}$. Diagonal products of representations for which $n-\bar n = 0$ define the vector $W^V_{1+\infty}$, diagonal products of representations for which $n+\bar n = 0$ define the axial-vector $W^A_{1+\infty}$. Let us introduce V and A integer labels
$l_{\rm V}= (n+\bar n)/2$ and $l_{\rm A}=(n-\bar n)/2$ and the vectors ${\bf l}=\left( l_{\rm V}, l_{\rm A}\right)$ and ${\bf t} = \left( 1, 1 \right)$. Then, the total charge number and spin (and statistics) of excitations can be expressed compactly as
\begin{eqnarray}
Q_{\rm V} + Q_{\rm A} &&= \ \ \ {\bf t}^T \cdot K^{-1} \cdot {\bf l} \ ,
\nonumber \\
S &&= {1\over 2} \ {\bf l}^T \cdot K^{-1} \cdot {\bf l} \ ,
\label{kmex}
\end{eqnarray}
where $Q_{\rm V} = Q$, $Q_{\rm A} = \Phi$ and $K^{-1}$ is the matrix
\begin{equation}
K^{-1} = 
\begin{bmatrix}
0 & 2(1+r)^2 \\
2(1+r)^2 & 0 \\
\end{bmatrix} \ .
\label{km}
\end{equation}
If we choose $r=-1/2$ and we restrict the excitation lattice to ${\mathbb Z}_2^2$ we obtain the $K$-matrix formulation of a ${\mathbb Z}_2$ spin liquid, with topological order degeneracy 4 \cite{wenmutual, lu1}; if we choose $2 (1+r)^2 = 1$, we obtain the $K$-matrix formulation of bosonic topological insulators \cite{dst}, an integer symmetry protected topological (SPT) phase \cite{lu2} which appears as a possible intermediate phase in the SIT, with charge and flux units $2e$ and $\pi/e$, and is also called a Bose metal because of the egde charge transport (for a review see \cite{bmreview}). Note that this phase is gapped when only matter fields are taken into account, the photon is massless in this phase, contrary to the superconductors and superinsualators to be discussed below. 
Both spin liquids and bosonic topological insulators admit more complex versions at $c=m>1$ with additional non-Abelian neutral (in both the V and A sectors) modes. 

Let us now discuss a purely diagonal $W^A_{1+\infty}$ minimal model with no charged excitations. From (\ref{qchiral}) and (\ref{qns}) we have $Q=(1+mr)^2 \sum_{i=1}^m (n_i + \bar n_i) $. Therefore, one can obtain such a purely axial minimal model by restricting to quantum numbers $( {\bf n}, {\bar {\bf n}})$, where $\bar n_1 = -n_m, \dots, \bar n_m = -n_1$, so that both the conditions $Q=0$ and $\bar n_1 \ge \bar n_2 \ge \dots \ge \bar n_m$ are satisfied. Using $\Phi=(1+mr)^2 \sum_{i=1}^m (n_i - \bar n_i) = 2(1+mr)^2 \sum_{i=1}^m n_i$ and $S=0$, we recognize that, for $2(1+mr)^2 = 1$, these models describe superconductors with spinless, integer vortices in units of $\Phi_0$, the vortex quantum. For $c=m=1$ these are the topological type-III superconductors realized near the SIT \cite{type31, type32}. For higher $c=m>1$ there are also here additional spinless neutral excitations carrying an $SU(m)$ isospin quantum number. 

Finally, let us address an example which shows that the $W_{1+\infty}$ infinite symmetry is a longer powerful constraint than topology. In superconductors, there are no charged excitations because the $U(1)$ charge is no more a good quantum number and only axial-vector excitations, vortices, survive. But there is another possibility to obtain models with no charged excitations, namely for the value $r=-1/m$, for which the lattice with metric (\ref{metric}) is degenerate and the charge ({\ref{qchiral}) is identically zero in both chiral sectors. Because the $\widehat U(1)$ factor decouples and only the neutral ${\cal W}_m$ factors survive , the effective central charge of these model is actually only $c=m-1$ and they do not admit a topological K-matrix formulation. 

Let us consider the simplest example $m=2$. Because of charge decoupling, these models are characterized by a unique half-integer label in each chiral sector: this is the $SU(2)$ isospin $i=(n_1-n_2)/2= k/2 \ge 0$, respectively $\bar i = (\bar n_1-\bar n_2)= \bar k/2 \ge 0$, where $n_{1,2}$ and $\bar n_{1,2}$ are the original integer labels $n_1 \ge n_2$, $\bar n_1 \ge \bar n_2$ and $k,\bar k \in {\mathbb N}$.  \cite{ctz1}. We shall again focus on models that can be written as ${\cal W}_2^V \otimes {\cal W}_2^A$ by restricting to a ${\mathbb N}^2$ sub-lattice for which $i+ \bar i = J \in {\mathbb N}$. Since $H= i^2+ \bar i^2$ and $S=i^2-\bar i^2= (i+\bar i) (i-\bar i)$, every axial diagonal in the admitted sublattice represents an $SO(3)$ spin multiplet with spin $\{ S \}_{\rm J} = J \{ -J, -J+1, \dots, 0, \dots, J-1, J \}$. 
Each state in this multiplet is characterized by its scaling dimension $H= i^2 + \bar i^2$. This is the meson spectrum of 2D superinsulators with strong electric interactions \cite{dst, vinokur, dtv}, the last possible state near the SIT. The first excitations are a scalar meson with $S=0$ and $H=1/2$ and a vector meson with $S=\pm 1$ and $H=1$. The former is the spinless electric pion formed by two Cooper pairs bound by a string of electric flux (for a review see \cite{enc}). 

\section{Conclusion}

We conclude by stressing that the infinite symmetry of 2D gapped quantum states involves both an infinite number of constraints and of quasi-local integrals of motion which imply insensitivity to weak disorder, this effects arising from strong interactions. In the case of superconductors, this is a generalization of Anderson's theorem due to strong magnetic interactions and is restricted only to the vortex sector, since the vector $W^V_{1+\infty}$ is entirely broken. In the case of bosonic topological insulators it manifests as a frozen state with no bulk conduction in both the charge and vortex sectors \cite{bmreview}. Finally, in the case of superinsulation, it is associated with charge confinement by electric interactions becoming so strong as to form electric flux tubes between charges \cite{enc}. Like conformal invariance is a much more powerful tool in 2D than in 3D (for a review see \cite{francesco}), also the presence of a gap has much more dramatic consequences in 2D than in 3D due to the character of the dynamical symmetry under quantum area-preserving diffeomorphisms.

\end{document}